\documentclass[sigconf,10pt]{acmart}

\usepackage{graphicx}
\usepackage{listings}
\usepackage{xspace}
\usepackage{xcolor}
\usepackage{listings}
\usepackage{subcaption}
\usepackage{enumitem}

\setcopyright{none}
\renewcommand\footnotetextcopyrightpermission[1]{}

\lstdefinestyle{customc}{
	belowcaptionskip=1\baselineskip,
	breaklines=true,
	linewidth=\columnwidth,
	columns=fullflexible,
	xleftmargin=0.05\columnwidth,
	xrightmargin=0.05\columnwidth,
	numbersep = -0.003\columnwidth,
	language=C,
	tabsize=2,
	showstringspaces=false,
	basicstyle=\footnotesize\ttfamily,
	keywordstyle=\bfseries\color{green!40!black},
	commentstyle=\itshape\color{purple!40!black},
	identifierstyle=\color{blue},
	stringstyle=\color{orange},
}

\newcommand{\ebpf}{eBPF\xspace}

\newcommand{\eg}{e.g.,\@\xspace}

\newcommand{\etc}{etc.\@\xspace}
\newcommand{\system}{\texttt{SandBPF}\xspace}
\newcommand{\noindgras}[1]{\noindent{\bf #1}}

\def\Snospace~{\S{}}

\begin{document}
	\title{Unleashing Unprivileged \ebpf Potential\\ with Dynamic Sandboxing}
	\author{Soo Yee Lim}
	\affiliation{%
		\institution{University of British Columbia}
		\city{Vancouver}
		\state{British Columbia}
		\country{Canada}
	}
\email{sooyee@cs.ubc.ca}
	\author{Xueyuan Han}
	\affiliation{%
		\institution{Wake Forest University}
		\city{Winston-Salem}
		\state{North Carolina}
		\country{USA}
	}
\email{vanbasm@wfu.edu}
	\author{Thomas Pasquier}
	\affiliation{%
		\institution{University of British Columbia}
		\city{Vancouver}
		\state{British Columbia}
		\country{Canada}
	}
\email{tfjmp@cs.ubc.ca}
	\date{}
	\renewcommand{\shortauthors}{Lim et al.}
	\renewcommand{\shorttitle}{Unleashing Unprivileged \ebpf Potential}

	\begin{abstract}
	
For safety reasons,
unprivileged users today have only limited ways
to customize the kernel through
the extended Berkeley Packet Filter (\ebpf).
This is unfortunate,
especially since
the \ebpf framework itself
has seen an increase in scope
over the years.
We propose \system, 
a software-based kernel isolation technique
that dynamically sandboxes \ebpf programs
to allow unprivileged users
to safely extend the kernel,
unleashing \ebpf's full potential.
Our early proof-of-concept
shows that \system 
can effectively prevent exploits
missed by \ebpf's native safety mechanism
(i.e., static verification)
while incurring 0\%-10\% overhead on web server benchmarks.

\noindgras{Note:} this a preprint version of the paper accepted at the 1st SIGCOMM Workshop on eBPF and Kernel Extensions~\cite{lim2023ebpf}.

	\end{abstract}



	\maketitle

	\section{Introduction}
	\label{sec:introduction}
	The extended Berkeley Packet Filter (\ebpf)
enables users to extend the Linux kernel's capabilities 
without modifying its source code.
To ensure safe extension of the kernel,
\ebpf uses a \emph{verifier}
to \emph{statically} verify the safety of an \ebpf program
before it is executed in the kernel.
Unfortunately, 
known vulnerabilities allow an \ebpf program 
to circumvent static verification checks, 
which enables a malicious \ebpf program 
to access arbitrary kernel memory~\cite{cve-2021-3490,cve-2021-33200,cve-2020-8835,cve-2021-31440} 
or execute arbitrary kernel code~\cite{cve-2021-29154}.

Most Linux distributions~\cite{unpriv_suse,unpriv_ubuntu}
err on the side of caution
by allowing \emph{only privileged users} 
to run \ebpf programs.
However, this restriction significantly limits the ability
for non-privileged applications
to customize the kernel.
For example,
it makes
adopting emerging \ebpf technologies
to support the implementation of specialized audit~\cite{lim2021secure},
scheduling~\cite{kaffes2021syrup},
and synchronization policies~\cite{park2022application}
in the kernel
for a particular application or container~difficult.

An alternative approach
is to formally verify that
the \ebpf verifier
\emph{correctly} guarantees 
the absence of \emph{all} possible attacks.
The \ebpf framework relies primarily on the verifier
to ensure the safety of an \ebpf program.
The verifier inspects
the program \emph{at load time},
so it imposes no run-time performance overhead. %
However,
prior incidents~\cite{stack_dos_cve,cve-2021-3490,cve-2021-33200,cve-2021-31440,cve-2022-23222,cve-2022-0264} have repeatedly shown
that a verified \ebpf program
is \emph{not always safe}.
The complexity of \ebpf programs
makes verifying these programs difficult,
resulting in both \emph{specification bugs}~\cite{stack_dos_cve}
(i.e., missing checks for a specific type of vulnerabilities)
and \emph{implementation bugs}~\cite{cve-2021-3490,cve-2021-33200,cve-2021-31440,cve-2022-23222,cve-2022-0264}
(i.e., incomplete checks for a supposedly checked vulnerability)
in the verifier.
As we elaborate in~\autoref{sec:motivation},
formally verifying the \ebpf verifier
cannot simultaneously resolve both types of bugs.

In light of these challenges,
we take a completely different approach
to enabling unprivileged \ebpf programs
to safely run in the Linux kernel.
We leverage software fault isolation (SFI)~\cite{wahbe1993efficient}, 
a software-based kernel isolation technique,
and binary rewriting
to \emph{dynamically sandbox}
an \ebpf program.
Our approach,
which we name \system, 
prevents an \ebpf program
from committing a memory safety violation
at run time
by confining all memory accesses to within the \ebpf sandbox
and limiting \ebpf control transfers to only valid call targets.
The proof-of-concept implementation
of \system shows that dynamic sandboxing is effective
in catching safety bugs that are missed by the verifier
while incurring reasonable performance overhead in realistic settings:
e.g., 0-10\% on web server macrobenchmarks.
These results are encouraging,
especially since we took a pure software approach
(as a first step to demonstrate efficacy),
rather than leveraging hardware support for isolation~\cite{pac,mte,intelvtx}
to improve performance, which we leave for future work.

\subsection*{Contributions.}
\begin{itemize}
	\item We study safety mechanisms in \ebpf 
        to motivate the need for \emph{dynamic isolation}
        (\autoref{sec:motivation}).
	\item We demonstrate that dynamic sandboxing is a viable solution to address \ebpf security concerns. 
        Our changes are a self-contained extension to the kernel and require no modification to existing workflows or programs (\autoref{sec:design}).
	\item We evaluate the performance overhead of our proof-of-concept
        implementation (\autoref{sec:evaluation}).
	\item We discuss the limitations of our implementation and 
        propose promising future research directions (\autoref{sec:discussion}).
\end{itemize}

\noindent \textbf{Disclaimer:} This work does not raise any ethical issues.

	\section{Motivation}
	\label{sec:motivation}
	\begin{figure}[!t]
	\centering
	\includegraphics[width=\columnwidth]{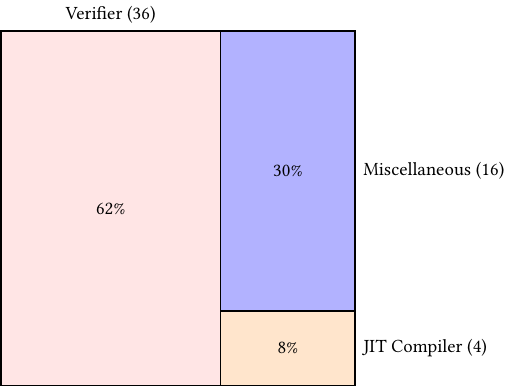}
	\caption{A tally of eBPF-related CVEs from 2010 to 2023. There are a total of 56 CVEs, the majority of which were discovered in the verifier.}
	\label{fig:cve}
\end{figure}

\begin{figure}[!t]
	
	\includegraphics[width=\columnwidth]{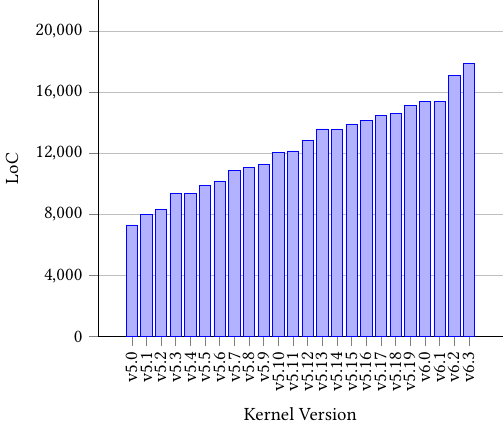}
	\caption{The evolution of the verifier's size in lines of code (LoC) from v5.0 in March 2019 (7,306 LoC) to v6.3 in April 2023 (17,904 LoC).}
\label{fig:verifier-size}
\end{figure}

The \ebpf verifier is the primary safety mechanism in \ebpf, 
but it has also been a major source of vulnerabilities (\autoref{fig:cve}).
Attackers can bypass the verifier and run malicious \ebpf programs %
by exploiting specification or implementation bugs in the verifier.
The verifier essentially works as a \emph{blocklist} of prohibited behaviors;
therefore, a specification bug exists
when a specific type of exploitable vulnerabilities 
is not considered in the blocklist.
For example,
early versions of the verifier neglected alignment checks for stack pointers,
which allowed adversaries to perform denial-of-service attacks~\cite{stack_dos_cve}.
Over the years,
the verifier has grown significantly
to mitigate specification bugs.
\autoref{fig:verifier-size} shows
that it has 
\emph{more than doubled} in the last four years.
Unfortunately, the ever-growing size and complexity %
makes it formidable to
formally verify the verifier in its entirety %
to eliminate any implementation bugs.
To the best of our knowledge,
no existing work has proved the \emph{completeness}
of the verifier's specification
or managed to formally verify
its current (likely still incomplete) implementation.
As a result, 
the Linux community has largely dismissed unprivileged \ebpf programs 
as an unsafe feature that should not be used~\cite{unprivebpf}, %
despite their great potential.

We propose to rely on a \emph{dynamic} enforcement mechanism,
by rewriting the binary code to insert run-time checks,
to \emph{protect memory accesses} and \emph{preserve control flow integrity}.
Note that 
the verifier performs checks beyond memory accesses and control flow integrity (\eg program termination).
Thus,
we still consider the verifier
to be an important safety component of \ebpf.
However,
our work can simplify the verifier's implementation (by removing memory access and control flow integrity checks),
thereby reducing its codebase and easing its formal verification efforts.
Our goal is to demonstrate the feasibility of dynamic enforcement;
therefore, 
we leave the simplification of the verifier to future work.

    \section{Threat Model}
    \label{sec:threat}
    We assume that an adversary can run unprivileged \ebpf programs. %
The adversary has no root access
and thus is unable to load kernel modules or modify kernel code.
However, they can exploit \ebpf vulnerabilities 
to gain arbitrary read or write access to kernel memory,
or execute arbitrary kernel code.
We assume a W$\oplus$X (write xor execute) enabled system,
so the adversary cannot overwrite any executable pages.

\begin{figure}[!t]
	\centering
	\includegraphics[width=\columnwidth]{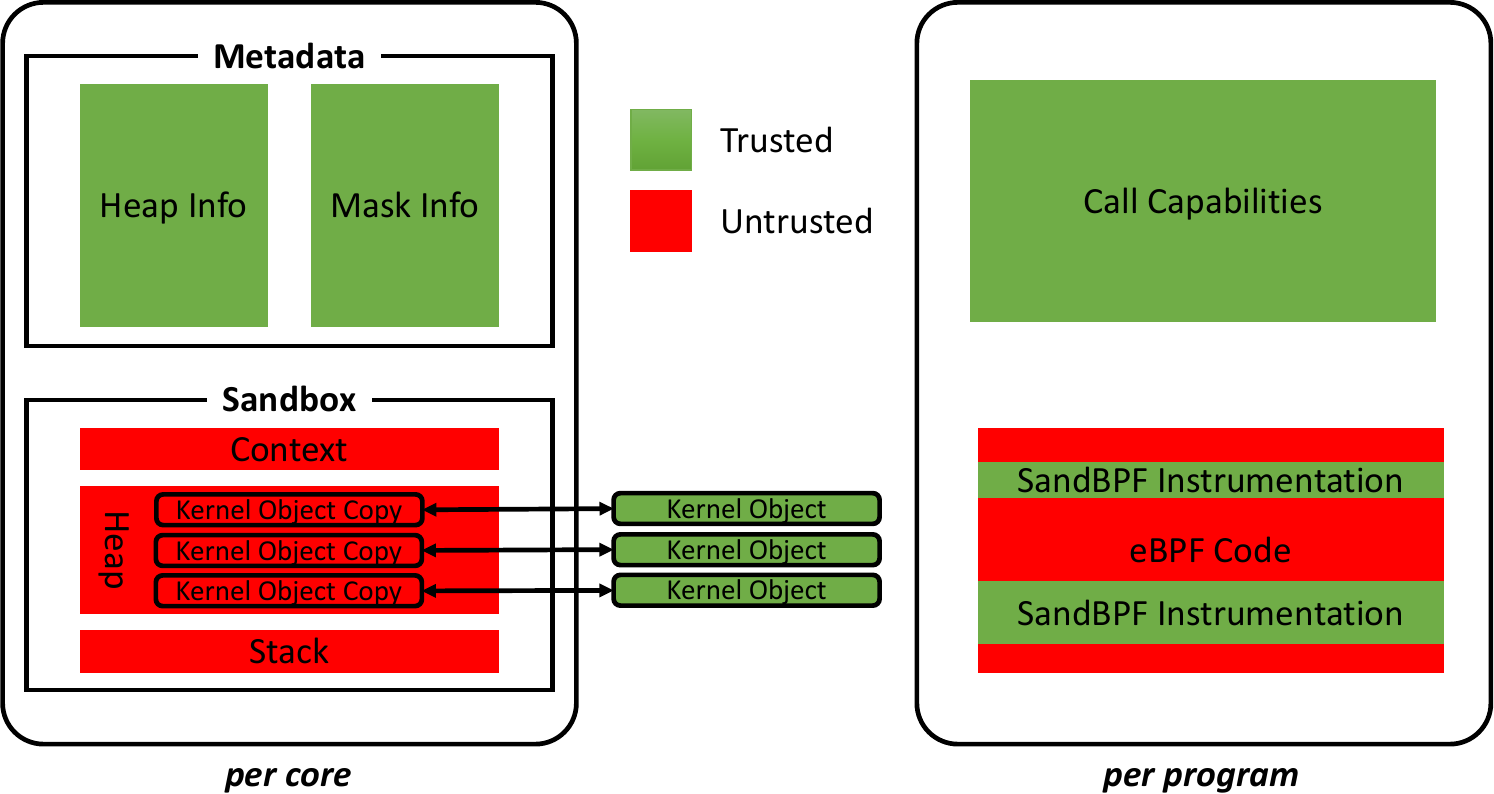}
	\caption{Illustration of the \system design and its (un)trusted components.}
	\label{fig:design}
\end{figure}

Our trusted computing base includes the OS kernel (excluding the \ebpf verifier and the JIT compiler)
and \system, whose correctness we plan to fully verify in future work.
As shown in~\autoref{fig:design}, 
both the data of an \ebpf program
that resides in the sandbox provided by \system
and the original \ebpf code are assumed to be untrusted. 
On the other hand, 
we assume \system's instrumentation
and its own data stored outside of the sandbox
(see details in~\autoref{sec:design})
to be trusted.
We do not consider attacks 
originated from anywhere else in the kernel 
except \ebpf.
Note that 
\system performs instrumentation 
on the final output of \ebpf's JIT compilation;
therefore,
it does not rely on
the correctness the JIT compiler or the verifier.
Like in prior kernel isolation work~\cite{swift2003improving, erlingsson2006xfi, ganapathy2007microdrivers, renzelmann2009decaf, castro2009fast, mao2011software,nikolaev2013virtuos,narayanan2019lxds,narayanan2020lightweight,mckee2022preventing}, 
side-channel attacks are orthogonal and thus out of scope.

	\section{Design \& Implementation}
	\label{sec:design}
	\begin{figure}[!t]
	\centering
	\includegraphics[width=\columnwidth]{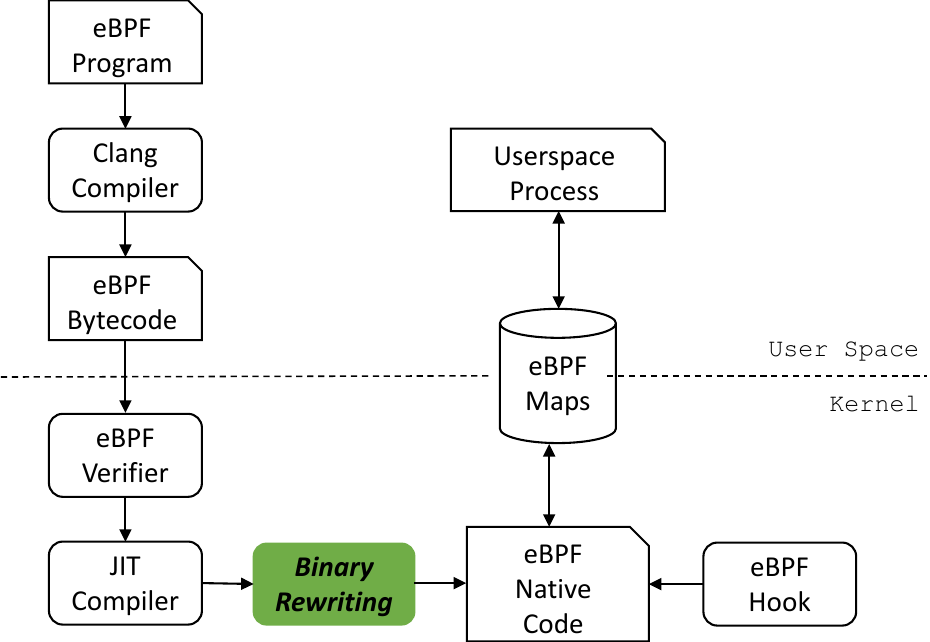}
	\caption{The workflow of a dynamically sandboxed \ebpf program. 
        \system's binary rewriting is trusted.}
	\label{fig:ebpf_sandbox_workflow}
\end{figure}

Keeping our security mechanism
completely transparent to \ebpf programmers
and ensuring compatibility with existing \ebpf programs
are the paramount design goals of \system.
As such,
\system is \emph{minimally invasive},
reusing the existing \ebpf pipeline and extending only what is necessary.
Specifically, \system adds binary rewriting only at the end of the JIT compilation, as shown in~\autoref{fig:ebpf_sandbox_workflow}. 
This allows experienced users to develop \ebpf programs like they normally do, 
while new users can rely on existing \ebpf documentation. 

Our \system proof-of-concept leverages software fault isolation (SFI)~\cite{wahbe1993efficient} 
to create a safe sandbox to execute \ebpf programs.
SFI is a software-based isolation technique 
that transforms memory-access and control-transfer instructions 
to prevent a program from accessing memory outside a designated region.
\autoref{fig:design} shows our SFI design.
\system ensures memory safety and control-flow integrity 
via \emph{address masking} and
\emph{trampolined control transfers}, respectively.
The former restricts
an \ebpf program's access
to only the memory within its address space,
and the latter
enforces that the program call only entry points on its allowlist
(based on its capabilities)
when jumping outside of its domain.
These checks together confine an \ebpf program 
to its own sandbox at run time. %

In the rest of this section,
we discuss \system's design in detail.
In~\autoref{sec:discussion}, 
we discuss the limitations of our current implementation.

\subsection{Memory Access Checks}
\label{sec:design:memory}

\system enforces memory safety by masking the target addresses of all read and write instructions, so that all memory accesses, including the out-of-bounds ones, always fall within the data region of an \ebpf domain. 
To avoid managing multiple address masks for data residing in different parts of the kernel address space, we reserve one memory page in each processor core to store the data of an \ebpf program, as shown in~\autoref{fig:design}. 
We disable interrupt and preemption during the execution of an \ebpf program, 
so each core runs only one \ebpf program at a time.

We emit an address masking check on \emph{every} read and write instruction. 
An address masking check consists of a bitwise-\texttt{and} instruction to clear the upper bits of the destination address, and subsequently a bitwise-\texttt{or} instruction to set the destination address to the memory region of an \ebpf sandbox.
For example, consider a 2048-byte aligned sandbox memory allocated at address \texttt{0xDEADB800}, and two pre-computed address masks (stored in the sandbox metadata as show in~\autoref{fig:design}): \texttt{and\_mask} (\texttt{0x7FF}) and \texttt{or\_mask} (\texttt{0xDEADB800}).
If an attacker attempts to perform an out-of-bounds memory access at \texttt{0xDEAF1234}, address masking would transform the target address to \texttt{0xDEADBA34}, which falls within the sandbox.
Thus, address masking guarantees that all memory accesses remain contained in the sandbox.

\subsection{Accessing Kernel Objects}
\label{sec:design:access}

An \ebpf program needs to access a kernel object
(1) when its input pointer (i.e., ``context'')
references a kernel object on program invocation,
or (2) when an \ebpf helper function
returns a pointer to a kernel data structure.
In the first case, we mirror the content of 
the data structure
in the context region of the sandbox
for the duration of the program execution. %
In the second case, 
we dynamically allocate heap space
in the sandbox to store a copy of the object.
As a concrete example,
consider the \ebpf ring buffer.
We modify the ring buffer's reserve/commit mechanism
to protect its access.
On \texttt{bpf\_ringbuf\_reserve}, we create a buffer on the sandbox heap corresponding to the reserved region.
Mapping information between the heap buffer and the reserved region 
is stored in the sandbox metadata,
which is not accessible by the \ebpf program.
On \texttt{bpf\_ringbuf\_commit}, the sandbox copy is synced with its corresponding kernel object.
This mechanism can be used for all similar operations.
We discuss some security-related issues in~\autoref{sec:discussion}.
 
\subsection{Control Flow Integrity (CFI)}
\label{sec:design:cfi}

We enforce CFI by redirecting all call instructions to a trampoline that checks the validity of the destination operand.
By design, \ebpf programs can interact with the kernel only through an allowlist of helper functions.
As different \ebpf program types have access to different sets of helper functions, we associate each \ebpf program type with a set of capabilities corresponding to the helper functions it is allowed to call.
In other words, the capabilities specify the valid entry points for control transfers in an \ebpf program, thereby preventing the program from executing arbitrary code in the kernel.
The capabilities are computed once for every program type at load time and stored in a hash table to provide an $O(1)$ search time.
\system %
dynamically checks if the \ebpf program has the capability to call the target.

	\section{Evaluation}
	\label{sec:evaluation}

We implemented \system for Linux 5.18.7.
All experiments were performed on a bare metal machine with 32GiB of RAM and an 8-core, 2.3GHz Intel Core i7 CPU. 
We disabled hyperthreading, turbo boost, and frequency scaling to reduce variance in performance benchmarking. 
We ran each experiment on two kernel configurations:
(1) The \emph{vanilla} configuration runs on the unmodified kernel, as our baseline.
(2) The \emph{sandbox} configuration runs on the same kernel but is instrumented with %
\system on \ebpf programs.

\subsection{Understanding %
Overhead}
\label{sec:evaluation:performance:breakdown}

\begin{table}[]
	\caption{The number of checks inserted and executed by \system
		in our example programs.}
	\label{tab:instrumentation}
	\resizebox{\columnwidth}{!}{%
		\begin{tabular}{|l|cc|cc|}
			\hline
			\multicolumn{1}{|c|}{} & \multicolumn{2}{|c|}{Injected} & \multicolumn{2}{c|}{Executed} \\ \cline{2-5} 
			\multicolumn{1}{|c|}{Program}                         & Address Masking  & Trampoline & Address Masking & Trampoline \\ \hline
			XDP           & 2   & 1  & 2     & 1   \\
			Socket Filter & 12  & 10 & 12    & 10  \\
			Katran   & 641 & 42 & 35-37 & 1-2 \\
			\hline
		\end{tabular}%
	}
\end{table}

\autoref{tab:instrumentation} shows the number of checks \system inserted 
into sandboxed \ebpf programs.
Note that the number of \emph{inserted} instrumentation points
does not directly influence performance;
rather, performance hinges on
the number of checks actually \emph{executed} at run time,
which in turn 
depends on execution paths.
In a complicated \ebpf program,
e.g., the Katran load balancer~\cite{katran},
we often observe fewer than $10$\% of the inserted checks
being executed at run time.

\begin{figure}[!t]
	\centering
	
	\includegraphics[width=\columnwidth]{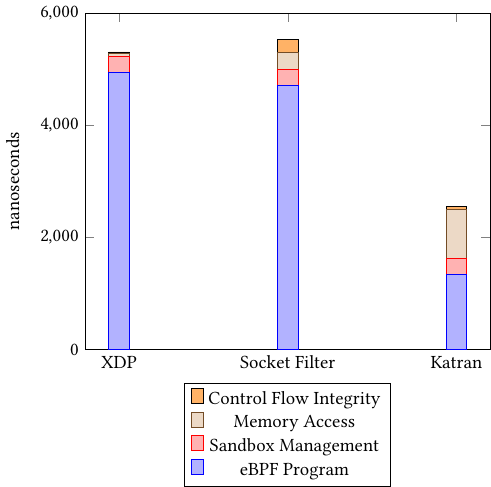}
	\caption{Breakdown of the overhead introduced by \system 
        on our three example \ebpf programs.}
	\label{fig:cost}
\end{figure}

In \autoref{fig:cost}, we 
run three \ebpf programs:
(1) The \texttt{XDP} program logs the size of each ingress packet entering a networking device.
(2) The \texttt{Socket Filter} program monitors packets by attaching itself to the 
\texttt{sock\_queue\_rcv\_skb()} function. 
It exchanges packet information with a userspace process through an \ebpf ring buffer.
(3) The \texttt{Katran} program performs load balancing and is attached to the NIC as an \texttt{xdp} program.
\texttt{Socket Filter} and \texttt{XDP} are programs 
provided as part of \texttt{libbpf-bootstrap}~\cite{libbpf-bootstrap},
a set of publicly available \ebpf example programs.
We decompose the overhead introduced by \system alongside three categories:

\noindgras{Sandbox Management}:
This corresponds to
sandbox initialization and shutdown
on the execution of an \ebpf program
(\eg preparing metadata, copying program parameters, 
switching execution context in and out of the sandbox, \etc).

\noindgras{Memory Access:}
 \system instruments both read and write instructions 
 to protect the confidentiality and integrity of kernel memory.
This differs from most SFI systems~\cite{wahbe1993efficient,erlingsson2006xfi,castro2009fast,erlingsson2006xfi,mao2011software} 
 that instrument only writes for performance. 

\noindgras{Control Flow Integrity:}
This overhead is due mostly to
the cost of linked list traversal
when \system searches through a hash table of call capabilities.
We plan to optimize \system's CFI checking performance
using a custom data structure.

We show the overhead in~\autoref{fig:cost}.
The overall overhead is a function of the number of checks executed in these programs:
\[
C_{overall} = C_{mem}(N_{mem}) + C_{tram}(N_{tram}) + C_{manage}
\]
where the overhead of memory access checks ($C_{mem}$) 
and that of CFI checks using the trampoline mechanism ($C_{tram}$)
are functions of the numbers of executed checks,
$N_{mem}$ and $N_{tram}$, respectively.
Both $N_{mem}$ and $N_{tram}$ correlate with the complexity of the programs themselves.
On the other hand,
the overhead of sandbox management ($C_{manage}$) is constant.

The \texttt{XDP} and \texttt{Socket Filter} programs perform tracing,
while \texttt{Katran} performs computations to make load-balancing decisions.
\texttt{XDP} and \texttt{Socket Filter} spend the majority of their time in helper functions, which are not instrumented, resulting in relatively low overhead.
On the other hand, \texttt{Katran} spend most of its time performing computation within instrumented code, resulting in a higher number of memory access checks and therefore a proportionally larger overhead.

\subsection{Microbenchmark}
\label{sec:evaluation:performance:micro}

\begin{table}[]
	\caption{Microbenchmark measuring the impact of \system on network communications running for 360s. (sf: send file, c$\rightarrow$s: client to server, s$\rightarrow$c: server to client).}
	\label{tab:netperf}
	\resizebox{\columnwidth}{!}{%
	\begin{tabular}{|lcccc|}
		\hline
		\multicolumn{1}{|c|}{} & \multicolumn{2}{|c|}{\textbf{XDP Program}} & \multicolumn{2}{c|}{\textbf{Socket Filter Program}} \\  
		\multicolumn{1}{|c|}{Test} & \multicolumn{1}{c}{Vanilla}  & \multicolumn{1}{c|}{\system} & \multicolumn{1}{c}{Vanilla} & \multicolumn{1}{c|}{\system} \\ \hline
		\multicolumn{5}{|c|}{Unidirectional throughput (MB/s)}                                                      \\ \hline
		\multicolumn{1}{|l|}{TCP sf}               & 40,588      & \multicolumn{1}{c|}{35,363 (\textbf{13\%})}          & 68,394                & 52,655 (\textbf{23\%})          \\ \hline
		\multicolumn{1}{|l|}{TCP c$\rightarrow$s}  & 36,740      & \multicolumn{1}{c|}{33,374 (\textbf{9\%})}           & 62,112                & 45,392 (\textbf{27\%})          \\ \hline
		\multicolumn{1}{|l|}{TCP s$\rightarrow$c}  & 36,626      & \multicolumn{1}{c|}{33,383 (\textbf{9\%})}           & 62,674                & 45,628 (\textbf{27\%})          \\ \hline
		\multicolumn{1}{|l|}{UDP s$\rightarrow$c}  & 48,019      & \multicolumn{1}{c|}{46,226 (\textbf{4\%})}           & 81,214                & 60,850 (\textbf{25\%})          \\ \hline
        \multicolumn{5}{|c|}{Round-trip transaction rate (transaction/s)}                                           \\ \hline
		\multicolumn{1}{|l|}{TCP}                  & 102,169     & \multicolumn{1}{c|}{86,416 (\textbf{15\%})}          & 135,713               & 94,611 (\textbf{30\%})          \\ \hline
		\multicolumn{1}{|l|}{UDP}                  & 118,409     & \multicolumn{1}{c|}{101,826 (\textbf{14\%})}         & 152,485               & 104,900 (\textbf{31\%})         \\ \hline
	\end{tabular}
}
\end{table}

We use netperf~\cite{netperf} to measure the overhead imposed by \system
on network communications. 
We run \texttt{XDP} and \texttt{Socket Filter}\footnote{
Due to space constraints,
we leave an extensive evaluation plan 
involving other \ebpf programs 
(which require more complex setups)
to future work.} on both kernel configurations
and measure the unidirectional throughputs 
and round-trip latencies for TCP and UDP.
We see in~\autoref{tab:netperf} that the overhead ranges from 4\% to 33\%.
We note that the high overhead
in the microbenchmark results
is largely the artifact
of netperf stressing the network interface
and therefore spending most of its execution time 
in kernel code that constantly triggers \ebpf events.
In practice, %
user applications typically perform meaningful computations in userspace, 
which would reduce the perceived \system overhead
as we show in~\autoref{sec:evaluation:performance:macro}.

\subsection{Macrobenchmark}
\label{sec:evaluation:performance:macro}

\begin{table}[]
	\caption{Macrobenchmark measuring web server performance of 20-1000 concurrent connections.}
	\resizebox{\columnwidth}{!}{%
	\begin{tabular}{|l|cc|cc|}
		\hline
		\multicolumn{1}{|c|}{} & \multicolumn{2}{|c|}{\textbf{XDP Program}} & \multicolumn{2}{c|}{\textbf{Socket Filter Program}} \\  
		\multicolumn{1}{|c|}{Test} & \multicolumn{1}{c}{Vanilla}  & \multicolumn{1}{c|}{\system} & \multicolumn{1}{c}{Vanilla} & \multicolumn{1}{c|}{\system} \\ \hline
		\multicolumn{5}{|c|}{Throughput (request/s)}                                                      \\
		\hline
		Apache 20                                  & 64,591                           & 64,526 (\textbf{0\%})                                 & 62,269                       & 60,089 (\textbf{4\%})                                \\
		Apache 100                                 & 86,190                           & 79,638 (\textbf{8\%})                                 & 87,576                       & 83,751 (\textbf{4\%})                                \\
		Apache 200                                 & 85,614                           & 81,749 (\textbf{5\%})                                 & 85,671                       & 83,381 (\textbf{3\%})                                \\
		Apache 500                                 & 68,329                           & 63,691 (\textbf{7\%})                                 & 72,399                       & 67,177 (\textbf{7\%})                                \\
		Apache 1000                                & 66,472                           & 62,508 (\textbf{6\%})                                 & 71,453                       & 66,171 (\textbf{7\%})                                \\ \hline
		Nginx 20                                   & 49,170                           & 45,731 (\textbf{7\%})                                 & 50,095                       & 45,331 (\textbf{10\%})                               \\
		Nginx 100                                  & 58,613                           & 54,494 (\textbf{7\%})                                 & 58,797                       & 54,029 (\textbf{8\%})                                \\
		Nginx 200                                  & 56,581                           & 53,051 (\textbf{6\%})                                 & 58,447                       & 53,869 (\textbf{8\%})                                \\
		Nginx 500                                  & 50,495                           & 47,699 (\textbf{6\%})                                 & 54,537                       & 50,822 (\textbf{7\%})                                \\
		Nginx 1000                                 & 46,302                           & 44,977 (\textbf{3\%})                                 & 50,651                       & 47,734 (\textbf{6\%})\\
		\hline                               
	\end{tabular}
}
\label{tab:macrobenchmark}

\end{table}

To evaluate the performance implication of \system
at the macro level,
we select a set of macrobenchmarks
(\emph{Apache} and \emph{Nginx})
from the Phoronix Test Suite~\cite{larabel2011phoronix}
that characterize whole-system performance
while stress-testing the network subsystems.
These web server benchmarks measure network throughputs,
which we report in~\autoref{tab:macrobenchmark}.
For all macrobenchmarks, 
we again run the \texttt{XDP} and \texttt{Socket Filter} programs 
to measure \system's overhead on these workloads.
We see that \system incurs no more than 10\%~overhead.

\subsection{Security}
\label{sec:evaluation:security}

\begin{table*}[]
	\centering
	\caption{\ebpf vulnerabilities that result in privilege escalation.}
\label{tab:exploit}
	\resizebox{\textwidth}{!}{
		\begin{tabular}{|l|p{5.5in}|}
			\hline
			\multicolumn{1}{|c|}{\textbf{CVE}} & \multicolumn{1}{c|}{\textbf{Vulnerability Description}} \\ \hline
			CVE-2021-3490 &
			The eBPF verifier incorrectly tracks the bounds of ALU32 bitwise operations, resulting in out-of-bounds reads and writes in the Linux kernel. \\ \cline{1-2} 
			CVE-2021-4204 & 
			The \ebpf verifier does not properly validate the bounds of \texttt{bpf\_ringbuf\_submit} and \texttt{bpf\_ringbuf\_discard} inputs, allowing out-of-bounds reads and writes in kernel memory. \\ \hline%
		\end{tabular}
	}
\end{table*}

We tested our proof-of-concept implementation against
two exploits that have publicly available source code
(\autoref{tab:exploit}).
These exploits leverage
\ebpf vulnerabilities
to violate the confidentiality and integrity of kernel memory.
Our experiments show that 
\system can successfully prevent CVE-2021-3490~\cite{cve-2021-3490}
and CVE-2021-4204~\cite{cve-2021-4204}.
For example,
in CVE-2021-3490,
a bounds-tracking bug in the \ebpf verifier
leads to out-of-bounds access.
An attacker can exploit
this vulnerability
to obtain arbitrary read and write accesses in the kernel memory.
Consequently,
they can leak \texttt{cred} pointers to userspace 
via \ebpf maps 
and escalate privilege by overwriting the \texttt{cred} structure.
To test both %
exploits, we ported \system to the affected Linux kernel version v5.8.0-25.26. 
\system successfully prevented both exploits
through address masking.
The attacker can no longer leak kernel pointers
to perform subsequent malicious activity (i.e., privilege escalation).

	\section{Discussion \& Future Work}
	\label{sec:discussion}
	\noindgras{Performance.}
Our primary focus is to demonstrate the tremendous potential 
of using dynamic sandboxing to improve \ebpf security
and validate our approach through a proof-of-concept implementation.
\system used stock kernel functions
(\eg the standard memory allocator \texttt{kmem\_alloc})
and data structures (\eg hash maps),
instead of any bespoke mechanisms
to optimize its run-time performance.
Moreover,
we made no use of asynchronous mechanisms, 
nor did we restrict \system's access checks 
to only \texttt{write} instructions. 
These ``tricks'' are often employed
in prior work to reduce the cost of SFI~\cite{wahbe1993efficient,erlingsson2006xfi,castro2009fast,erlingsson2006xfi,mao2011software}.
Therefore,
one could consider
our current proof-of-concept implementation to be 
a \emph{worst-case scenario} in terms of performance.
Even so,
we see only $\leq$10\% overhead
from \system while providing fully-fledged
memory and control flow protection
on macrobenchmarks (\autoref{sec:evaluation:performance:macro}).
This overhead is a reasonable baseline
for our future work to improve performance.
For example,
hardware features,
such as ARM's Pointer Authentication Code (PAC)~\cite{arm_pac} 
and Memory Tagging Extension (MTE)~\cite{mte},
are promising avenues to expedite memory access checks (which constitute most of the overhead in non-tracing tasks as discussed in \autoref{sec:evaluation:performance:breakdown}).
We emphasize that at the moment,
the \emph{only} alternative to our approach
is to entirely disable unprivileged \ebpf programs.

\noindgras{Security.}
We backported \system to earlier versions of the kernel 
to demonstrate its ability to safeguard vulnerabilities 
in the verifier (\autoref{sec:evaluation:security}).
While %
\system's binary rewriting approach
can reduce the kernel's attack surface,
we recognize that our current evaluation is insufficient,
given the innate complexity of 
proving the effectiveness of a sandboxing technique~\cite{besson2019compiling}.
We are in the process of designing 
a more extensive evaluation methodology based on \emph{fault injection}.
We note that 
to fully unleash the potential of unprivileged \ebpf programs,
one must consider security issues beyond
the ones addressed by \system through dynamic sandboxing;
vulnerabilities stemming from
\eg kernel namespacing and shared resources~\cite{gao2019houdini,yang2021demons,liu2023kit}
will also need to be tackled.
Furthermore,
the verifier currently restricts
an \ebpf program's ability to modify certain attributes
of some kernel objects (e.g., \texttt{sk\_buff}).
We plan to dynamically enforce such an restriction
in the syncing mechanism described in~\autoref{sec:design:access}.

\noindgras{Towards Simplifying the Verifier.}
Our approach makes it possible
to replace a subset of the verifier's 
compilation-time checks,
such as memory accesses 
(which are difficult and sometimes unsound %
to perform statically~\cite{cve-2021-3490,cve-2021-33200,cve-2021-31440,cve-2022-23222,cve-2022-0264}),
with \system's dynamic checks.
This not only simplifies the verifier
by obviating the need to check aspects of correctness
that have been proven hard to guarantee,
but more importantly,
allows the \ebpf framework
to relax some constraints
imposed on its programs
(\eg in dynamic memory allocation).
These constraints exist
due to
the difficulties of verifying statically
the safety of an \ebpf program.
There is a wealth of opportunities to explore 
ways to relax \ebpf constraints
to enrich its functionality,
which we leave to future work.

	\section{Related Work}
	\label{sec:related}
	\system leverages a software-based %
isolation technique,
specifically SFI,
to dynamically sandbox \ebpf programs. 
SFI instruments %
code with dynamic checks 
to ensure that run-time data accesses and control flow transfers 
are within specified bounds.
Prior work~\cite{small1998misfit,erlingsson2006xfi,castro2009fast,mao2011software} 
leveraged SFI to sandbox OS extensions such as device drivers.
For example, BGI~\cite{castro2009fast} associates an access control list 
with each byte of memory 
to specify byte-level access permissions. %
\system instead enforces SFI at the page level, 
restricting the access of a sandboxed \ebpf program 
to %
pre-defined pages of kernel memory.
Due to performance concerns,
most SFI systems~\cite{erlingsson2006xfi,mao2011software},
including BGI, do not check read instructions;
as a result, they cannot provide \emph{confidentiality} guarantees.
In contrast,
to account for potential leakage of kernel memory to userspace,
\system instruments both read and write instructions, 
thus assuring both confidentiality and integrity of kernel memory.
This design decision comes at the cost of performance as discussed in~\autoref{sec:discussion}.

Other than SFI,
prior work~\cite{zhou2006safedrive} also proposed
to use implicit pointer bounds information
to enforce fine-grained type and memory safety.
For example,
SafeDrive~\cite{zhou2006safedrive} allows
developers to provide type annotations
that describe pointer bounds
to insert dynamic checks in device drivers.
Unlike SafeDrive,
\system instrumentation is completely automated,
requiring no manual annotation effort.
This is possible thanks to the relatively well-defined \ebpf API
and a well-scoped set of kernel objects
that an \ebpf program is usually allowed to interact with,
as compared to Linux kernel modules or device drivers.

Recent work~\cite{jia2023kernel} has also proposed a new Rust-based \ebpf design. 
It leverages the Rust tool-chain to perform static checks 
(\eg memory safety and control-flow integrity).
In addition,
it uses run-time mechanisms to enforce properties 
such as program termination
that Rust does not natively provide.
While this approach eliminates the need 
for an in-kernel \ebpf verifier,
it has two main drawbacks.
First, 
the Rust verification tool-chain must be executed 
by a trusted third party before signing the \ebpf programs.
As a result,
this approach limits the kernel to load 
only \ebpf programs %
that are signed by trusted third parties,
as the kernel itself can no longer independently verify them.
This runs contrary to \system's philosophy 
of increasing \ebpf usage as an end goal.
Second, vulnerabilities in the complex Rust ecosystem~\cite{cve_rust} take us back to the same problem with the \ebpf verifier -- static analysis alone is insufficient to guarantee run-time safety of \ebpf programs. 
In other words, \ebpf extensions can exploit Rust verifier's vulnerabilities to corrupt kernel memory at run time.
Therefore, we believe that dynamic sandboxing is a key step towards unleashing the potential of unprivileged \ebpf.

	\section{Conclusion}
	\label{sec:conclusion}
	We show that dynamic sandboxing is a viable 
approach to enforce a number of security properties
in \ebpf programs,
complementary to the current static mechanism
employed by the \ebpf verifier.
Dynamic sandboxing will not replace verification;
instead, it enhances run-time safety of the kernel
to justify the (currently dismissed) support of unprivileged \ebpf programs.
\system, our proof-of-concept implementation based on software fault isolation, 
incurs reasonable overhead.
We believe that our work opens up an interesting design space,
which allows future work to bring competitive performance improvements to this approach, 
particularly by leveraging available hardware features.

	\section*{Acknowledgment}
	We acknowledge the support of the Natural Sciences and Engineering Research Council of Canada (NSERC). Nous remercions le Conseil de
recherches en sciences naturelles et en génie du Canada (CRSNG) de son soutien.
This material is based upon work supported 
by the U.S. National Science Foundation under Grant CNS-2245442.
Any opinions, findings, and conclusions or recommendations 
expressed in this material are those of the author(s) 
and do not necessarily reflect the views of the National Science Foundation.

	\bibliographystyle{ACM-Reference-Format}
	\bibliography{bibliography}


\begin{thebibliography}{43}


\ifx \showCODEN    \undefined \def \showCODEN     #1{\unskip}     \fi
\ifx \showDOI      \undefined \def \showDOI       #1{#1}\fi
\ifx \showISBNx    \undefined \def \showISBNx     #1{\unskip}     \fi
\ifx \showISBNxiii \undefined \def \showISBNxiii  #1{\unskip}     \fi
\ifx \showISSN     \undefined \def \showISSN      #1{\unskip}     \fi
\ifx \showLCCN     \undefined \def \showLCCN      #1{\unskip}     \fi
\ifx \shownote     \undefined \def \shownote      #1{#1}          \fi
\ifx \showarticletitle \undefined \def \showarticletitle #1{#1}   \fi
\ifx \showURL      \undefined \def \showURL       {\relax}        \fi
\providecommand\bibfield[2]{#2}
\providecommand\bibinfo[2]{#2}
\providecommand\natexlab[1]{#1}
\providecommand\showeprint[2][]{arXiv:#2}

\bibitem[\protect\citeauthoryear{??}{arm}{[n. d.]}]%
        {arm_pac}
 \bibinfo{year}{[n. d.]}\natexlab{}.
\newblock \bibinfo{title}{{ARMv8.3 Pointer Authentication}}.
\newblock   (\bibinfo{year}{[n. d.]}).
\newblock
\showURL{%
\url{https://lwn.net/Articles/718888/}}


\bibitem[\protect\citeauthoryear{??}{mte}{[n. d.]}]%
        {mte}
 \bibinfo{year}{[n. d.]}\natexlab{}.
\newblock \bibinfo{title}{{ARMv8.5-A Memory Tagging Extension}}.
\newblock \bibinfo{howpublished}{Online (Accessed: \today)}.
  (\bibinfo{year}{[n. d.]}).
\newblock
\newblock
\shownote{\url{https://developer.arm.com/documentation/102925/0100}.}


\bibitem[\protect\citeauthoryear{??}{pac}{[n. d.]}]%
        {pac}
 \bibinfo{year}{[n. d.]}\natexlab{}.
\newblock \bibinfo{title}{{ARMv8.5-A Pointer Authentication}}.
\newblock \bibinfo{howpublished}{Online (Accessed: \today)}.
  (\bibinfo{year}{[n. d.]}).
\newblock
\newblock
\shownote{\url{shorturl.at/GHM69}.}


\bibitem[\protect\citeauthoryear{??}{sta}{[n. d.]}]%
        {stack_dos_cve}
 \bibinfo{year}{[n. d.]}\natexlab{}.
\newblock \bibinfo{title}{{CVE-2017-17856}}.
\newblock \bibinfo{howpublished}{Online (Accessed: \today)}.
  (\bibinfo{year}{[n. d.]}).
\newblock
\newblock
\shownote{\url{https://cve.mitre.org/cgi-bin/cvename.cgi?name=CVE-2017-17856}.}


\bibitem[\protect\citeauthoryear{??}{cve}{[n. d.]a}]%
        {cve-2020-8835}
 \bibinfo{year}{[n. d.]}\natexlab{a}.
\newblock \bibinfo{title}{{CVE-2020-8835}}.
\newblock \bibinfo{howpublished}{Online (Accessed: \today)}.
  (\bibinfo{year}{[n. d.]}).
\newblock
\newblock
\shownote{\url{https://cve.mitre.org/cgi-bin/cvename.cgi?name=CVE-2020-8835}.}


\bibitem[\protect\citeauthoryear{??}{cve}{[n. d.]b}]%
        {cve-2021-29154}
 \bibinfo{year}{[n. d.]}\natexlab{b}.
\newblock \bibinfo{title}{{CVE-2021-29154}}.
\newblock \bibinfo{howpublished}{Online (Accessed: \today)}.
  (\bibinfo{year}{[n. d.]}).
\newblock
\newblock
\shownote{\url{https://cve.mitre.org/cgi-bin/cvename.cgi?name=CVE-2021-29154}.}


\bibitem[\protect\citeauthoryear{??}{cve}{[n. d.]c}]%
        {cve-2021-31440}
 \bibinfo{year}{[n. d.]}\natexlab{c}.
\newblock \bibinfo{title}{{CVE-2021-31440}}.
\newblock \bibinfo{howpublished}{Online (Accessed: \today)}.
  (\bibinfo{year}{[n. d.]}).
\newblock
\newblock
\shownote{\url{https://cve.mitre.org/cgi-bin/cvename.cgi?name=CVE-2021-31440}.}


\bibitem[\protect\citeauthoryear{??}{cve}{[n. d.]d}]%
        {cve-2021-33200}
 \bibinfo{year}{[n. d.]}\natexlab{d}.
\newblock \bibinfo{title}{{CVE-2021-33200}}.
\newblock \bibinfo{howpublished}{Online (Accessed: \today)}.
  (\bibinfo{year}{[n. d.]}).
\newblock
\newblock
\shownote{\url{https://cve.mitre.org/cgi-bin/cvename.cgi?name=CVE-2021-33200}.}


\bibitem[\protect\citeauthoryear{??}{cve}{[n. d.]e}]%
        {cve-2021-3490}
 \bibinfo{year}{[n. d.]}\natexlab{e}.
\newblock \bibinfo{title}{{CVE-2021-3490}}.
\newblock \bibinfo{howpublished}{Online (Accessed: \today)}.
  (\bibinfo{year}{[n. d.]}).
\newblock
\newblock
\shownote{\url{https://cve.mitre.org/cgi-bin/cvename.cgi?name=CVE-2021-3490}.}


\bibitem[\protect\citeauthoryear{??}{cve}{[n. d.]f}]%
        {cve-2021-4204}
 \bibinfo{year}{[n. d.]}\natexlab{f}.
\newblock \bibinfo{title}{{CVE-2021-4204}}.
\newblock \bibinfo{howpublished}{Online (Accessed: \today)}.
  (\bibinfo{year}{[n. d.]}).
\newblock
\newblock
\shownote{\url{https://cve.mitre.org/cgi-bin/cvename.cgi?name=CVE-2021-4204}.}


\bibitem[\protect\citeauthoryear{??}{cve}{[n. d.]g}]%
        {cve-2022-0264}
 \bibinfo{year}{[n. d.]}\natexlab{g}.
\newblock \bibinfo{title}{{CVE-2022-0264}}.
\newblock \bibinfo{howpublished}{Online (Accessed: \today)}.
  (\bibinfo{year}{[n. d.]}).
\newblock
\newblock
\shownote{\url{https://cve.mitre.org/cgi-bin/cvename.cgi?name=CVE-2022-0264}.}


\bibitem[\protect\citeauthoryear{??}{cve}{[n. d.]h}]%
        {cve-2022-23222}
 \bibinfo{year}{[n. d.]}\natexlab{h}.
\newblock \bibinfo{title}{{CVE-2022-23222}}.
\newblock \bibinfo{howpublished}{Online (Accessed: \today)}.
  (\bibinfo{year}{[n. d.]}).
\newblock
\newblock
\shownote{\url{https://cve.mitre.org/cgi-bin/cvename.cgi?name=CVE-2022-23222}.}


\bibitem[\protect\citeauthoryear{??}{lib}{[n. d.]}]%
        {libbpf-bootstrap}
 \bibinfo{year}{[n. d.]}\natexlab{}.
\newblock \bibinfo{title}{ibbpf-bootstrap}.
\newblock   (\bibinfo{year}{[n. d.]}).
\newblock


\bibitem[\protect\citeauthoryear{??}{cve}{[n. d.]i}]%
        {cve_rust}
 \bibinfo{year}{[n. d.]}\natexlab{i}.
\newblock \bibinfo{title}{{Mitre: Rust CVEs}}.
\newblock \bibinfo{howpublished}{Online (Accessed: \today)}.
  (\bibinfo{year}{[n. d.]}).
\newblock
\newblock
\shownote{\url{https://cve.mitre.org/cgi-bin/cvekey.cgi?keyword=rust}.}


\bibitem[\protect\citeauthoryear{??}{net}{[n. d.]}]%
        {netperf}
 \bibinfo{year}{[n. d.]}\natexlab{}.
\newblock \bibinfo{title}{Netperf}.
\newblock \bibinfo{howpublished}{Online (Accessed: \today)}.
  (\bibinfo{year}{[n. d.]}).
\newblock
\newblock
\shownote{\url{https://hewlettpackard.github.io/netperf/}.}


\bibitem[\protect\citeauthoryear{??}{unp}{[n. d.]a}]%
        {unprivebpf}
 \bibinfo{year}{[n. d.]}\natexlab{a}.
\newblock \bibinfo{title}{{Reconsidering unprivileged BPF}}.
\newblock \bibinfo{howpublished}{Online (Accessed: \today)}.
  (\bibinfo{year}{[n. d.]}).
\newblock
\newblock
\shownote{\url{https://lwn.net/Articles/796328/}.}


\bibitem[\protect\citeauthoryear{??}{unp}{[n. d.]b}]%
        {unpriv_suse}
 \bibinfo{year}{[n. d.]}\natexlab{b}.
\newblock \bibinfo{title}{{Security Hardening: Use of eBPF by unprivileged
  users has been disabled by default}}.
\newblock \bibinfo{howpublished}{Online (Accessed: \today)}.
  (\bibinfo{year}{[n. d.]}).
\newblock
\newblock
\shownote{\url{https://www.suse.com/support/kb/doc/?id=000020545}.}


\bibitem[\protect\citeauthoryear{??}{unp}{[n. d.]c}]%
        {unpriv_ubuntu}
 \bibinfo{year}{[n. d.]}\natexlab{c}.
\newblock \bibinfo{title}{{Unprivileged eBPF disabled by default for Ubuntu
  20.04 LTS, 18.04 LTS, 16.04 ESM}}.
\newblock \bibinfo{howpublished}{Online (Accessed: \today)}.
  (\bibinfo{year}{[n. d.]}).
\newblock
\newblock
\shownote{\url{https://discourse.ubuntu.com/t/27047}.}


\bibitem[\protect\citeauthoryear{Besson, Blazy, Dang, Jensen, and Wilke}{Besson
  et~al\mbox{.}}{2019}]%
        {besson2019compiling}
\bibfield{author}{\bibinfo{person}{Fr{\'e}d{\'e}ric Besson},
  \bibinfo{person}{Sandrine Blazy}, \bibinfo{person}{Alexandre Dang},
  \bibinfo{person}{Thomas Jensen}, {and} \bibinfo{person}{Pierre Wilke}.}
  \bibinfo{year}{2019}\natexlab{}.
\newblock \showarticletitle{Compiling sandboxes: Formally verified software
  fault isolation}. In \bibinfo{booktitle}{{\em European Symposium on
  Programming (ESOP'19)}}. Springer, \bibinfo{pages}{499--524}.
\newblock


\bibitem[\protect\citeauthoryear{Castro, Costa, Martin, Peinado, Akritidis,
  Donnelly, Barham, and Black}{Castro et~al\mbox{.}}{2009}]%
        {castro2009fast}
\bibfield{author}{\bibinfo{person}{Miguel Castro}, \bibinfo{person}{Manuel
  Costa}, \bibinfo{person}{Jean-Philippe Martin}, \bibinfo{person}{Marcus
  Peinado}, \bibinfo{person}{Periklis Akritidis}, \bibinfo{person}{Austin
  Donnelly}, \bibinfo{person}{Paul Barham}, {and} \bibinfo{person}{Richard
  Black}.} \bibinfo{year}{2009}\natexlab{}.
\newblock \showarticletitle{Fast byte-granularity software fault isolation}. In
  \bibinfo{booktitle}{{\em Symposium on Operating Systems Principles
  (SOSP'09)}}. ACM, \bibinfo{pages}{45--58}.
\newblock


\bibitem[\protect\citeauthoryear{Erlingsson, Abadi, Vrable, Budiu, and
  Necula}{Erlingsson et~al\mbox{.}}{2006}]%
        {erlingsson2006xfi}
\bibfield{author}{\bibinfo{person}{Ulfar Erlingsson},
  \bibinfo{person}{Mart{\'\i}n Abadi}, \bibinfo{person}{Michael Vrable},
  \bibinfo{person}{Mihai Budiu}, {and} \bibinfo{person}{George~C Necula}.}
  \bibinfo{year}{2006}\natexlab{}.
\newblock \showarticletitle{XFI: Software guards for system address spaces}. In
  \bibinfo{booktitle}{{\em Symposium on Operating Systems Design and
  Implementation (OSDI'06)}}. USENIX, \bibinfo{pages}{75--88}.
\newblock


\bibitem[\protect\citeauthoryear{Ganapathy, Balakrishnan, Swift, and
  Jha}{Ganapathy et~al\mbox{.}}{2007}]%
        {ganapathy2007microdrivers}
\bibfield{author}{\bibinfo{person}{Vinod Ganapathy}, \bibinfo{person}{Arini
  Balakrishnan}, \bibinfo{person}{Michael~M Swift}, {and}
  \bibinfo{person}{Somesh Jha}.} \bibinfo{year}{2007}\natexlab{}.
\newblock \showarticletitle{Microdrivers: A new architecture for device
  drivers}.
\newblock \bibinfo{journal}{{\em Network\/}}  \bibinfo{volume}{134}
  (\bibinfo{year}{2007}), \bibinfo{pages}{27--8}.
\newblock


\bibitem[\protect\citeauthoryear{Gao, Gu, Li, Jamjoom, and Wang}{Gao
  et~al\mbox{.}}{2019}]%
        {gao2019houdini}
\bibfield{author}{\bibinfo{person}{Xing Gao}, \bibinfo{person}{Zhongshu Gu},
  \bibinfo{person}{Zhengfa Li}, \bibinfo{person}{Hani Jamjoom}, {and}
  \bibinfo{person}{Cong Wang}.} \bibinfo{year}{2019}\natexlab{}.
\newblock \showarticletitle{Houdini's escape: Breaking the resource rein of
  linux control groups}. In \bibinfo{booktitle}{{\em Conference on Computer and
  Communications Security (CCS'19)}}. ACM, \bibinfo{pages}{1073--1086}.
\newblock


\bibitem[\protect\citeauthoryear{Incubator}{Incubator}{[n. d.]}]%
        {katran}
\bibfield{author}{\bibinfo{person}{Meta Incubator}.} \bibinfo{year}{[n.
  d.]}\natexlab{}.
\newblock \bibinfo{title}{{Katran:A high performance layer 4 load balancer}}.
\newblock \bibinfo{howpublished}{Online (Accessed: \today)}.
  (\bibinfo{year}{[n. d.]}).
\newblock
\newblock
\shownote{\url{https://github.com/facebookincubator/katran}.}


\bibitem[\protect\citeauthoryear{Jia, Sahu, Oswald, Williams, Le, and Xu}{Jia
  et~al\mbox{.}}{2023}]%
        {jia2023kernel}
\bibfield{author}{\bibinfo{person}{Jinghao Jia}, \bibinfo{person}{Raj Sahu},
  \bibinfo{person}{Adam Oswald}, \bibinfo{person}{Dan Williams},
  \bibinfo{person}{Michael~V Le}, {and} \bibinfo{person}{Tianyin Xu}.}
  \bibinfo{year}{2023}\natexlab{}.
\newblock \showarticletitle{{Kernel Extension Verification is Untenable}}. In
  \bibinfo{booktitle}{{\em Workshop on Hot Topics in Operating Systems
  (HotOS'23)}}. ACM, \bibinfo{pages}{150--157}.
\newblock


\bibitem[\protect\citeauthoryear{Kaffes, Humphries, Mazi{\`e}res, and
  Kozyrakis}{Kaffes et~al\mbox{.}}{2021}]%
        {kaffes2021syrup}
\bibfield{author}{\bibinfo{person}{Kostis Kaffes}, \bibinfo{person}{Jack~Tigar
  Humphries}, \bibinfo{person}{David Mazi{\`e}res}, {and}
  \bibinfo{person}{Christos Kozyrakis}.} \bibinfo{year}{2021}\natexlab{}.
\newblock \showarticletitle{Syrup: User-defined scheduling across the stack}.
  In \bibinfo{booktitle}{{\em Symposium on Operating Systems Principles
  (SOSP'21)}}. ACM, \bibinfo{pages}{605--620}.
\newblock


\bibitem[\protect\citeauthoryear{Larabel and Tippett}{Larabel and
  Tippett}{2011}]%
        {larabel2011phoronix}
\bibfield{author}{\bibinfo{person}{Michael Larabel} {and}
  \bibinfo{person}{Matthew Tippett}.} \bibinfo{year}{2011}\natexlab{}.
\newblock \showarticletitle{Phoronix test suite}.
\newblock \bibinfo{journal}{{\em Phoronix Media,[Online]. Available:
  http://www. phoronix-test-suite. com/.[Accessed June 2016]\/}}
  (\bibinfo{year}{2011}).
\newblock


\bibitem[\protect\citeauthoryear{Lim, Han, and Pasquier}{Lim
  et~al\mbox{.}}{2023}]%
        {lim2023ebpf}
\bibfield{author}{\bibinfo{person}{Soo~Yee Lim}, \bibinfo{person}{Xueyuan Han},
  {and} \bibinfo{person}{Thomas Pasquier}.} \bibinfo{year}{2023}\natexlab{}.
\newblock \showarticletitle{{Unleashing Unprivileged eBPF Potential with
  Dynamic Sandboxing}}. In \bibinfo{booktitle}{{\em SIGCOMM Workshop on eBPF
  and Kernel Extensions}}. ACM.
\newblock


\bibitem[\protect\citeauthoryear{Lim, Stelea, Han, and Pasquier}{Lim
  et~al\mbox{.}}{2021}]%
        {lim2021secure}
\bibfield{author}{\bibinfo{person}{Soo~Yee Lim}, \bibinfo{person}{Bogdan
  Stelea}, \bibinfo{person}{Xueyuan Han}, {and} \bibinfo{person}{Thomas
  Pasquier}.} \bibinfo{year}{2021}\natexlab{}.
\newblock \showarticletitle{Secure Namespaced Kernel Audit for Containers}. In
  \bibinfo{booktitle}{{\em Symposium on Cloud Computing (SoCC'21)}}. ACM,
  \bibinfo{pages}{518--532}.
\newblock


\bibitem[\protect\citeauthoryear{Liu, Gong, and Fonseca}{Liu
  et~al\mbox{.}}{2023}]%
        {liu2023kit}
\bibfield{author}{\bibinfo{person}{Congyu Liu}, \bibinfo{person}{Sishuai Gong},
  {and} \bibinfo{person}{Pedro Fonseca}.} \bibinfo{year}{2023}\natexlab{}.
\newblock \showarticletitle{KIT: Testing OS-Level Virtualization for Functional
  Interference Bugs}. In \bibinfo{booktitle}{{\em International Conference on
  Architectural Support for Programming Languages and Operating Systems
  (ASPLOS'23)}}. ACM, \bibinfo{pages}{427--441}.
\newblock


\bibitem[\protect\citeauthoryear{Mao, Chen, Zhou, Wang, Zeldovich, and
  Kaashoek}{Mao et~al\mbox{.}}{2011}]%
        {mao2011software}
\bibfield{author}{\bibinfo{person}{Yandong Mao}, \bibinfo{person}{Haogang
  Chen}, \bibinfo{person}{Dong Zhou}, \bibinfo{person}{Xi Wang},
  \bibinfo{person}{Nickolai Zeldovich}, {and} \bibinfo{person}{M~Frans
  Kaashoek}.} \bibinfo{year}{2011}\natexlab{}.
\newblock \showarticletitle{Software fault isolation with API integrity and
  multi-principal modules}. In \bibinfo{booktitle}{{\em Symposium on Operating
  Systems Principles (SOSP'11)}}. ACM, \bibinfo{pages}{115--128}.
\newblock


\bibitem[\protect\citeauthoryear{McKee, Giannaris, Perez, Shrobe, Payer,
  Okhravi, and Burow}{McKee et~al\mbox{.}}{2022}]%
        {mckee2022preventing}
\bibfield{author}{\bibinfo{person}{Derrick McKee}, \bibinfo{person}{Yianni
  Giannaris}, \bibinfo{person}{Carolina~Ortega Perez}, \bibinfo{person}{Howard
  Shrobe}, \bibinfo{person}{Mathias Payer}, \bibinfo{person}{Hamed Okhravi},
  {and} \bibinfo{person}{Nathan Burow}.} \bibinfo{year}{2022}\natexlab{}.
\newblock \showarticletitle{{Preventing Kernel Hacks with HAKC}}. In
  \bibinfo{booktitle}{{\em Network and Distributed System Security Symposium
  (NDSS'22)}}. Internet Society.
\newblock


\bibitem[\protect\citeauthoryear{Narayanan, Balasubramanian, Jacobsen, Spall,
  Bauer, Quigley, Hussain, Younis, Shen, Bhattacharyya,
  et~al\mbox{.}}{Narayanan et~al\mbox{.}}{2019}]%
        {narayanan2019lxds}
\bibfield{author}{\bibinfo{person}{Vikram Narayanan}, \bibinfo{person}{Abhiram
  Balasubramanian}, \bibinfo{person}{Charlie Jacobsen}, \bibinfo{person}{Sarah
  Spall}, \bibinfo{person}{Scott Bauer}, \bibinfo{person}{Michael Quigley},
  \bibinfo{person}{Aftab Hussain}, \bibinfo{person}{Abdullah Younis},
  \bibinfo{person}{Junjie Shen}, \bibinfo{person}{Moinak Bhattacharyya},
  {et~al\mbox{.}}} \bibinfo{year}{2019}\natexlab{}.
\newblock \showarticletitle{{LXDs: Towards Isolation of Kernel Subsystems}}. In
  \bibinfo{booktitle}{{\em Annual Technical Conference (ATC'19)}}. USENIX,
  \bibinfo{pages}{269--284}.
\newblock


\bibitem[\protect\citeauthoryear{Narayanan, Huang, Tan, Jaeger, and
  Burtsev}{Narayanan et~al\mbox{.}}{2020}]%
        {narayanan2020lightweight}
\bibfield{author}{\bibinfo{person}{Vikram Narayanan}, \bibinfo{person}{Yongzhe
  Huang}, \bibinfo{person}{Gang Tan}, \bibinfo{person}{Trent Jaeger}, {and}
  \bibinfo{person}{Anton Burtsev}.} \bibinfo{year}{2020}\natexlab{}.
\newblock \showarticletitle{{Lightweight kernel isolation with virtualization
  and VM functions}}. In \bibinfo{booktitle}{{\em International Conference on
  Virtual Execution Environments}}. ACM, \bibinfo{pages}{157--171}.
\newblock


\bibitem[\protect\citeauthoryear{Neiger, Santoni, Leung, Rodgers, and
  Uhlig}{Neiger et~al\mbox{.}}{2006}]%
        {intelvtx}
\bibfield{author}{\bibinfo{person}{Gil Neiger}, \bibinfo{person}{Amy Santoni},
  \bibinfo{person}{Felix Leung}, \bibinfo{person}{Dion Rodgers}, {and}
  \bibinfo{person}{Rich Uhlig}.} \bibinfo{year}{2006}\natexlab{}.
\newblock \showarticletitle{Intel virtualization technology: Hardware support
  for efficient processor virtualization.}
\newblock \bibinfo{journal}{{\em Intel Technology Journal\/}}
  \bibinfo{volume}{10}, \bibinfo{number}{3} (\bibinfo{year}{2006}).
\newblock


\bibitem[\protect\citeauthoryear{Nikolaev and Back}{Nikolaev and Back}{2013}]%
        {nikolaev2013virtuos}
\bibfield{author}{\bibinfo{person}{Ruslan Nikolaev} {and}
  \bibinfo{person}{Godmar Back}.} \bibinfo{year}{2013}\natexlab{}.
\newblock \showarticletitle{VirtuOS: An operating system with kernel
  virtualization}. In \bibinfo{booktitle}{{\em Symposium on Operating Systems
  Principles (SOSP'13)}}. ACM, \bibinfo{pages}{116--132}.
\newblock


\bibitem[\protect\citeauthoryear{Park, Zhou, Qian, Calciu, Kim, and
  Kashyap}{Park et~al\mbox{.}}{2022}]%
        {park2022application}
\bibfield{author}{\bibinfo{person}{Sujin Park}, \bibinfo{person}{Diyu Zhou},
  \bibinfo{person}{Yuchen Qian}, \bibinfo{person}{Irina Calciu},
  \bibinfo{person}{Taesoo Kim}, {and} \bibinfo{person}{Sanidhya Kashyap}.}
  \bibinfo{year}{2022}\natexlab{}.
\newblock \showarticletitle{{Application-Informed Kernel Synchronization
  Primitives}}. In \bibinfo{booktitle}{{\em Symposium on Operating Systems
  Design and Implementation (OSDI'22)}}. USENIX, \bibinfo{pages}{667--682}.
\newblock


\bibitem[\protect\citeauthoryear{Renzelmann and Swift}{Renzelmann and
  Swift}{2009}]%
        {renzelmann2009decaf}
\bibfield{author}{\bibinfo{person}{Matthew~J Renzelmann} {and}
  \bibinfo{person}{Michael~M Swift}.} \bibinfo{year}{2009}\natexlab{}.
\newblock \showarticletitle{Decaf: Moving Device Drivers to a Modern
  Language.}. In \bibinfo{booktitle}{{\em Annual Technical Conference
  (ATC'09)}}. USENIX.
\newblock


\bibitem[\protect\citeauthoryear{Small and Seltzer}{Small and Seltzer}{1998}]%
        {small1998misfit}
\bibfield{author}{\bibinfo{person}{Christopher Small} {and}
  \bibinfo{person}{Margo Seltzer}.} \bibinfo{year}{1998}\natexlab{}.
\newblock \showarticletitle{MiSFIT: Constructing safe extensible systems}.
\newblock \bibinfo{journal}{{\em IEEE concurrency\/}} \bibinfo{volume}{6},
  \bibinfo{number}{3} (\bibinfo{year}{1998}), \bibinfo{pages}{34--41}.
\newblock


\bibitem[\protect\citeauthoryear{Swift, Bershad, and Levy}{Swift
  et~al\mbox{.}}{2003}]%
        {swift2003improving}
\bibfield{author}{\bibinfo{person}{Michael~M Swift}, \bibinfo{person}{Brian~N
  Bershad}, {and} \bibinfo{person}{Henry~M Levy}.}
  \bibinfo{year}{2003}\natexlab{}.
\newblock \showarticletitle{Improving the reliability of commodity operating
  systems}. In \bibinfo{booktitle}{{\em Symposium on Operating Systems
  Principles (SOSP'03)}}. ACM, \bibinfo{pages}{207--222}.
\newblock


\bibitem[\protect\citeauthoryear{Wahbe, Lucco, Anderson, and Graham}{Wahbe
  et~al\mbox{.}}{1993}]%
        {wahbe1993efficient}
\bibfield{author}{\bibinfo{person}{Robert Wahbe}, \bibinfo{person}{Steven
  Lucco}, \bibinfo{person}{Thomas~E Anderson}, {and} \bibinfo{person}{Susan~L
  Graham}.} \bibinfo{year}{1993}\natexlab{}.
\newblock \showarticletitle{{Efficient Software-based Fault Isolation}}. In
  \bibinfo{booktitle}{{\em Symposium on Operating Systems Principles
  (SOSP'93)}}. ACM, \bibinfo{pages}{203--216}.
\newblock


\bibitem[\protect\citeauthoryear{Yang, Shen, Li, Yang, Lu, Xiao, Zhou, Qin, Yu,
  Ma, et~al\mbox{.}}{Yang et~al\mbox{.}}{2021}]%
        {yang2021demons}
\bibfield{author}{\bibinfo{person}{Nanzi Yang}, \bibinfo{person}{Wenbo Shen},
  \bibinfo{person}{Jinku Li}, \bibinfo{person}{Yutian Yang},
  \bibinfo{person}{Kangjie Lu}, \bibinfo{person}{Jietao Xiao},
  \bibinfo{person}{Tianyu Zhou}, \bibinfo{person}{Chenggang Qin},
  \bibinfo{person}{Wang Yu}, \bibinfo{person}{Jianfeng Ma}, {et~al\mbox{.}}}
  \bibinfo{year}{2021}\natexlab{}.
\newblock \showarticletitle{Demons in the shared kernel: Abstract resource
  attacks against os-level virtualization}. In \bibinfo{booktitle}{{\em
  Conference on Computer and Communications Security (CCS'21)}}. ACM,
  \bibinfo{pages}{764--778}.
\newblock


\bibitem[\protect\citeauthoryear{Zhou, Condit, Anderson, Bagrak, Ennals,
  Harren, Necula, and Brewer}{Zhou et~al\mbox{.}}{2006}]%
        {zhou2006safedrive}
\bibfield{author}{\bibinfo{person}{Feng Zhou}, \bibinfo{person}{Jeremy Condit},
  \bibinfo{person}{Zachary Anderson}, \bibinfo{person}{Ilya Bagrak},
  \bibinfo{person}{Rob Ennals}, \bibinfo{person}{Matthew Harren},
  \bibinfo{person}{George Necula}, {and} \bibinfo{person}{Eric Brewer}.}
  \bibinfo{year}{2006}\natexlab{}.
\newblock \showarticletitle{SafeDrive: Safe and recoverable extensions using
  language-based techniques}. In \bibinfo{booktitle}{{\em Symposium on
  Operating Systems Design and Implementation (OSDI'06)}}. USENIX,
  \bibinfo{pages}{45--60}.
\newblock


\end{thebibliography}

\end{document}